\documentstyle[preprint,aps]{revtex}
\begin{document}
\preprint{Usach/96/05, IFUSP/P-1218}

\title{GLOBAL ANOMALIES AND ANYONS IN 1+1 DIMENSIONS}
\author{J. GAMBOA$^1$\thanks{E-mail: jgamboa@lauca.usach.cl },  
V. O. RIVELLES$^2$\thanks{E-mail: vrivelles@if.usp.br} and 
J. ZANELLI$^{1,3}$\thanks{E-mail: jz@einstein.cecs.cl}}
\address{$^1$ Departamento de F\'{\i}sica, Universidad de Santiago 
de Chile, 
Casilla 307, Santiago 2, Chile 
\\ $^2$ Instituto de F\'\i sica, Universidade de S\~ao Paulo, 
C.P. 66318, 05389-970 S\~ao Paulo, Brazil
\\ $^3$ Centro de Estudios Cient\'\i ficos de Santiago, Casilla
16443, Santiago 9, Chile}
\date{\today}
\maketitle 

\begin{abstract} 
We consider the analog in one spatial dimension of the
Bose-Fermi transmutation for planar systems. A quantum mechanical
system of a spin 1/2 particle coupled to an abelian gauge field,
which is classically invariant under gauge transformations and
charge conjugation is studied. It is found that unless the flux 
enclosed by the particle orbits is quantized, and the spin takes a value
$n+ 1/2$, at least one of the two symmetries would be anomalous.
Thus, charge conjugation invariance and the existence of abelian
instantons simultaneously force the particles to be either
bosons or fermions, but not anyons.

\end{abstract}
\narrowtext

It is the aim of theoretical physics to provide economical
explanations for the fundamental features of nature. Two
brilliant examples of this are the quantization of spin in
multiples of $\hbar/2$, and the connection between spin and the
(anti-) symmetry of the wave function under exchange of
identical particles. It has been powerfully stressed however,
that both results are crucially dependent on the fact that the
rotation group in three spatial dimensions is non-abelian. Since
this fails to be so for lower dimensions, spin is not
necessarily quantized and the states need not be symmetric or 
antisymmetric under particle exchange.

In the last ten years intense research on the physical and
mathematical properties of planar --{\it i.e.},
(2+1)-dimensional-- systems has taken place. From the
mathematical point of view, the discovery of topological quantum
field theories \cite{witten} and fractional statistics
\cite{wilczek} are surely the most important results. On
the other hand, the fact that spin and statistics could be
fractional and their potential applications is, probably, one
the most exciting discoveries in theoretical physics.

However although there are many well established theoretical
\cite{mattis} and experimental results\cite{fisher} for linear
--{\it i.e.}, (1+1)-dimensional-- systems, a deeper
understanding of their origin is still lacking. 

One of these intriguing results is the bosonization in
one-dimensional (1+1) systems. It is generally 
believed that bosonization occurs naturally in those systems
because there is no rotation group in one dimension and, as a
consequence, the spin could be considered a matter of
convention. 

There are two approaches to bosonization: A non-linear,
non-local field transformation that maps a fermionic action into
a bosonic one \cite{JW,mande,cole}; and the more recent
two-dimensional construction
of Polyakov \cite{poli}, where the fermions are integrated out
and the resulting effective action is written in terms of purely
bosonic fields . These two approaches seem to be related, as it
can be argued in the context of duality \cite{burgess}.

In all approaches, topology plays a key role. The integer spin
excitations are found to exist in a topologically non-trivial
sector of the bosonized theory. The resulting (effective) action
has a boundary term of topological origin that is responsible
for the antisymmetry of the wave function under particle
exchange, thus respecting the fermionic statistics in the \lq\lq
bosonized" theory.

The construction based on the non-local mapping, the
appearence of this topological boundary term is not explicit
because the identification is either made at the classical
Lagrangian level \cite{mande}, or it is perturbative
\cite{cole}.  Its is therefore of interest to investigate a
simple system where the topological features of the mapping can
be studied in the quantum theory.

The approach based on the integration over the fermionic
fields can be exemplified with a relativitic spinning particle
in two spatial dimensions. Upon integration of the spin degrees
of freedom in the action $S[x,\theta]$, the resulting effective
action reads \cite{dhar}, 
\begin{equation} 
S_{eff} = \int d\tau [ m\sqrt{{\dot x}^2} + {1\over 2}{\cal W}],
\label{action} 
\end{equation} 
which is just a bosonic way of describing a relativistic
particle of spin one-half. The writhing number, ${\cal W}$, is a
topological invariant which classically does not contribute to
the equations of motion but quantum mechanically is responsible
for the non-trivial character of the Bose-Fermi transmutation.
The factor ${1\over 2}$ is precisely the spin of the particle.

One could also note that (\ref{action}) is a
particular case of a more general construction. In fact, when
one couples a spinless particle to an abelian Chern-Simons field
$A_{\mu}$ in a planar system, the action reads\cite{zee} 
\begin{equation} 
S = \int d\tau [ m\sqrt{{\dot x}^2} + A_\mu {\dot x}^\mu] 
+ {1\over 2\sigma} \int d^3x \epsilon^{\mu\nu\rho} A_\mu \partial_\nu
A_\rho.  
\end{equation} 
Integrating out the gauge field, one finds the effective Lagrangian
\begin{equation}
L_{eff} = m\sqrt{{\dot x}^2} + {\sigma\over 4\pi}{\cal W},
\end{equation}
which matches (\ref{action}) for $\sigma = 2\pi$. The
coefficient ${\sigma\over 4\pi}$ corresponds to the spin of the
effective system and, in this sense, (3) describes a quantum
particle with fractional spin and statistics \cite{forte}. 
In one spatial dimension the rotation group is discrete; its
representations are one-dimensional and labeled by a phase. 
This phase
(spin) can in turn  be shown to be determined by the class of
boundary conditions that render the Hamiltonian self-adjoint
\cite{gz}. If no further constraints are imposed on this phase,
the spin can take any real value, interpolating continuously
between bosons and fermions. If one takes  this idea seriously,
then a quantum field theory in 1+1 dimensions should be
anyonizable and not just bosonizable, as is commonly assumed. 

The purpose of this letter is to show how it is possible
to write an expression analogous to (3) for a simple quantum
mechanical system in one dimension, and to discuss some
subtleties associated with the bosonization procedure.  In 
particular, we will show how, after bosonization, the system
remembers that it comes from a fermionic one.  

 
In order to describe our results let us proceed in analogy with
analysis of \cite{poli} and \cite{dhar}, starting with a
non-relativistic spinning particle described by the action 

\begin{equation} 
S = \int_{t_1}^{t_2} dt \biggl[{1\over 2}{\dot x}^2 -{1\over 2} V^2 (x) + 
\psi^\dagger ( i\partial_t + A ) \psi \biggr], \label{lagra} 
\end{equation} 
which, for $A=V^{\prime}$ is $N=1$ supersymmetric quantum
mechanics (SSQM), where $V(x)$ is the superpotential. 

This action (\ref{lagra}) has two classical symmetries {\it v.i.z.} 

i) Invariance under local $U(1)$ gauge transformations  
\begin{equation} 
\psi^{'} (t) = e^{i\omega (t) }\psi (t), \,\,\,\,\,\,\,\,\,\,\,\,\,\,\,\,
\psi^{'\dagger} (t) = e^{-i\omega (t) }\psi^{\dagger} (t),
\label{gauge} 
\end{equation} 
where the gauge potential $A$ transform as \cite{gauge}  
\begin{equation} 
A \rightarrow A + \frac{d\omega}{dt}. 
\label{gauge1}
\end{equation} 

ii) Invariance under \lq \lq charge-conjugation" {\it i.e.}  
\begin{equation} 
\psi \leftrightarrow
\psi^\dagger,\,\,\,\,\,\,\,\,\,\,\,\,\,\,\,\, 
A \rightarrow - A. \label{cc}
\end{equation}  

The partition function is defined as 
\begin{equation} 
Z = \int {\cal D} x {\cal D} \psi {\cal D} \psi^\dagger 
e^{-S}.
\label{propa} 
\end{equation} 

It is customary to assume the orbits to be periodic in the bosonic
coordinates, while either periodic \cite{windey} or antiperiodic
\cite{ssqm} in the fermions. Since we want to investigate the
possibility of having anyons, which should not be expected to
obey neither periodic nor antiperiodic boundary conditions, we
will allow for twisted boundary conditions, namely 
\begin{equation}
x(t_1)=x(t_2),\label{gbc1}
\end{equation}
and
\begin{equation} 
\psi (t_2) = e^{2\pi i\alpha} \psi (t_1), \label{gbc2}
\end{equation} 
where $\alpha $ is an arbitrary real number. 

Integrating  over the fermionic variables, 
\begin{equation} 
Z_{\alpha} = \int {\cal D} x \det ( i\partial_t + A )_{\alpha} 
e^{i\int_{t_1}^{t_2} dt ({1\over 2} {\dot x}^2 - {1\over 2} V^2 )}.
\label{propa1} 
\end{equation} 

As usual, the determinant of an operator $\Omega$ is computed as
$\det (\Omega)_{\alpha} = \prod_n \lambda^{(\alpha)}_n$, where
$\lambda^{(\alpha)}_n$ are the eigenvalues \cite{dunne}. Using (\ref{gbc2}),
the eigenvalues are 
\begin{equation} 
\lambda^{(\alpha)}_n = {1\over T} \int_{t_1}^{t_2} dt A (x) - 
{{2\pi (\alpha + n)}\over T}, \label{eigen}
\end{equation} 
where $T = t_2 - t_1$, and the fermionic determinant becomes  
\begin{eqnarray} 
\Gamma_\alpha (A) &=& \det ( i\partial_t + A )_{\alpha}
\nonumber \\ 
&=& \prod_{n=-\infty}^{n=\infty} 
\biggl[ {1\over T} \int_{t_1}^{t_2} dt A (x) + 
{{2\pi (\alpha + n)}\over T} \biggr]. \label{gamma}
\end{eqnarray}

In order to compute the infinite product, one can isolate the
$n=0$ eigenvalue, then standard manipulations lead to

\begin{equation} 
\Gamma_\alpha (A) = {1\over T} (y + 2\pi \alpha)
\prod_{n=1}^{\infty} {{({-2\pi \over T})}^2 n^2}  
\prod_{n=1}^{\infty} \biggl[ 1 - {{{({y + 2\pi}) \alpha}}^2\over 4n^2 }
\biggr], 
\end{equation} 
with $y = \int_{t_1}^{t_2} dt A (x)$ and, using well known identities, we
arrive at \cite{footnote4}
 
\begin{equation} 
\Gamma_\alpha (A ) = {\cal N} \sin \biggl
[\int_{t_1}^{t_2} dt \biggl({1\over 2} A + 
\frac{\pi \alpha}{T} \biggr)\biggr], \label{deter1} 
\end{equation} 
where ${\cal N}$ is a normalization constant independent of $x$. 

This formula reduces to the known results when $\alpha =0$
(bosons), and $\alpha =1/2$ (fermions). This can be checked by
direct calculation in SSQM \cite{ssqm}. 

The determinant (\ref{deter1}) can be understood as follows.
Assume one starts with some definite boundary condition for the
fermions, say periodic boundary conditions. Then eq.
(\ref{deter1}) can be viewed as the result of a gauge
transformation on the gauge potential $A$ of the form
(\ref{gauge1}) with $\omega$ given by 
\begin{equation} 
\omega (t)= {2\pi \alpha \over {t_2 - t_1}} {(t - t_1)}.
\label{fixed} 
\end{equation}
As a consequence, there is a one to one correspondence between
gauge transformations of the class (\ref{fixed}) and the twist
chosen for the fermionic boundary condition (\ref{gbc2}). Thus,
by means of successive gauge transformations of this kind 
one can continuosly
interpolate between $\Gamma_{\alpha=0} (A)$ (bosons) and
$\Gamma_{\alpha=1/2} (A)$ (fermions).  

Thus, the complete partition function is   

\begin{equation} 
Z_{\alpha} = \int {\cal D} x \,\,\, \Gamma_\alpha (A) 
\displaystyle{
e}^{ i\int_{t_1}^{t_2} dt({1\over 2} {\dot x}^2 - {1\over 2}
V^2)}.  \label{propa2}
\end{equation}

At this point one can ask whether the classical symmetries of
the model are respected in the quantum theory. In order to
address this question, one notes that gauge invariance requires
$\Gamma_\alpha (A) = \Gamma_0 (A)$, and therefore either
\begin{equation}
a) \,\,\,\,\,\,\,\,\,\,\,\,\,\,\,\, \alpha = 2n, \label{a} 
\end{equation}
or
\begin{equation}
b) \,\,\,\,\,\,\,\,\,\,\,\,\,\,\,\, \Phi =(2n+1-\alpha)\pi,
\label{b} 
\end{equation}
where $\Phi\equiv \int_0^T dt A$ is the flux enclosed by the
particle orbit in 
Euclidean space (instanton). On the other hand, invariance under
charge conjugation implies $\Gamma_\alpha (A) = \Gamma_\alpha
(-A)$ and hence,
\begin{equation} 
c) \,\,\,\,\,\,\,\,\,\,\,\,\,\,\,\, \alpha = n+1/2, \label{c} 
\end{equation}
or
\begin{equation}
d) \,\,\,\,\,\,\,\,\,\,\,\,\,\,\,\, \Phi\equiv \int_0^T dt A =2n
\pi. \label{d} 
\end{equation}

It is clear from this that if the flux $\underline{were\,\, not}$ quantized,
it would be impossible to respect both symmetries
simultaneously. Furthermore, if $\alpha \epsilon {\bf Z}$, then
\begin{equation}
\Gamma_\alpha (A) = (-1)^n sin[\int_0^T dt A],
\label{Z}
\end{equation}
is an odd functional of $A$, which combined with charge
conjugation invariance implies that $\Gamma_\alpha (A) = 0$, and
the theory would be inconsistent. Thus, the only combination of
the above conditions that ensures consistency and absence of
anomalies is $b$ and $c$, namely,

\begin{equation}
\Phi =(m+1/2)\pi, \,\,\, \,\,\, \alpha= n+1/2, \,\,\, \,\,\, n-m
= \mbox{odd}.
\end{equation} 

Thus, we see that 
\begin{equation} 
\Gamma_\alpha (A ) = {\cal N^{'}}(\alpha) \displaystyle {e}^{
i\int_{t_1}^{t_2} dt A[x(t)]},
\label{deter} 
\end{equation} 
where ${\cal N^{'}}(\alpha)$ is a normalization constant, and
the effective action (17) is
\begin{equation} 
S = \int_{t_1}^{t_2} dt [{1\over 2} {\dot x}^2 - {1\over 2} V^2 + {1\over 2} A].
\label{final}
\end{equation} 

In conclusion, (\ref{final}) is the two-dimensional analog of
(3), where the flux ${1\over 2}\int_{t_1}^{t_2} dt A = {1\over 2} \Phi$ plays
the role of ${\sigma\over 4\pi} {\cal W}$ in 2+1 dimensions.
However, this analogy is not correct unless the flux is
quantized, which occurs for the cases $3$ and $4$ as shown in
the the following table.  
\vspace{1.0cm}

\begin{tabular}{|c|lr|c|c|c|}	
\hline
       & $\Phi$  &  $\alpha$ & Gauge Inv. & Ch. Conj. & Anomalous \\ \hline
1\,\,\,& arbitrary &   $2n$      & yes & no  & yes \\ \hline
2\,\,\, & arbitrary &  $n+1/2$   & no  & yes & yes \\ \hline
3\,\,\, & $2n\pi$  &  noninteger & no  & yes & yes \\ \hline
4\,\,\, & $(m+1/2)\pi$ &  $n+1/2$  & yes & yes & no  \\
 \,\,\, &              & $(n-m)$ odd&     &     &     \\ \hline
\end{tabular} 
\vspace{1.0cm}

Possibility $3$, however does not preserve gauge invariance and
the theory is anomalous. Case $4$ is more interesting because
it shows that the two classical symmetries are preserved and
$\alpha$ takes on half-integer values only. This means that the
particles described by the bosonized action (\ref{final}) are not
bosons --as naively expected--, nor anyons --as the elementary
group-theoretical analysis suggests--, but in fact
$\underline{fermions}$. 

In the path integral (or partition function) it is customary to
integrate over periodic (antiperiodic) orbits for bosonic
(fermionic) variables. The reason for this is essentially
classical: it is under these conditions that the action has
an extremum on the classical orbits \cite{ht}. Nevertheless,
it is not obvious that this is necessarily so at the quantum
level. What one learns from the preceding analysis is that
unless fermions are antiperiodic, the theory would not respect
gauge and charge conjugation invariance.  

This result can also be reached through a geometrical analysis.
The boundary conditions (\ref{gbc1}) and (\ref{gbc2}) correspond
to superimposing a gauge transformation on the orbit so that the
spinor comes back to itself, modulo a finite gauge
transformation, when $x$ completes a full turn.  This implies
that the family of boundary conditions considered splits into the
homotopy classes in $\Pi_1({SO(2)/ U(1)}) \simeq \Pi_1(U(1))=$
{\bf Z}. These classes are labeled by a value of $\alpha \in
${\bf Z}, or $\alpha \in${\bf Z}$+1/2$ \cite{goddard}. The
additional requirement of charge conjugation invariance, rules
out the integer values for $\alpha$.

The flux quantization results from the compactification of the
time direction as a consequence of the (anti-) periodicity of
the fields. This is analogous to the quantization of the abelian
Chern-Simons coefficient in 2+1 dimensions when the theory is
defined on a multiply connected manifold \cite{nino}. The fact
that the flux enclosed by the orbits is quantized can also be
interpreted as a condition for quantization of the orbits,
similar to the Bohr-Sommerfeld rule.

The extension of these results to higher dimensions and their
connection with non abelian anomalies will discussed elsewhere
\cite{jvj}. 

Several enlightening discusions with N. Brali$\acute{c}$ are warmly
acknowledged. This work was partially supported by grants
1950278 and 1960229 by FONDECYT-Chile, grants 04-953/ZI and
04-953/GR from DICYT-USACH. V.O.R. was partially supported from
CNPq-Brazil. J.Z. wishes to also acknowledge partial support by
a group of chilean private companies (COPEC, CMPC, CGE, MINERA
ESCONDIDA, NOVAGAS, BUSINESS DESIGN ASS. and XEROX-Chile). One
of us (J.G) is a recipient of a John S. Guggenheim Fellowship.


\begin{references}
\bibitem{gauge} Here one should note that although the gauge
field is not dynamical, our analysis goes through unchanged if
one adds the term $\frac{1}{4}\int d^2x F^2$ to the action.
\bibitem{footnote4} Of course, one could try other
regularization procedures. For instance, one can factorize $x$ in
the product $\prod_n (x + n)$ instead of $n$. However, this 
regularization it is not convenient here.
\end{references}
\end{document}